\documentclass[preprint,showpacs,showkeys,aps]{revtex4}

\usepackage{graphicx}

\begin{document}
                      
\title{                 
Hamiltonian Dynamics of Thermostated Systems:                \\
Two-Temperature Heat-Conducting $\phi^4$ Chains.
}

\author{
Wm. G. Hoover and
Carol G. Hoover                                              \\
Highway Contract 60, Box 565                                 \\
Ruby Valley, Nevada 89833  and                               \\  
HiTech Center, Great Basin College                           \\
Elko, Nevada 89801                                              
}

\date{\today}



\begin{abstract}

We consider and compare four Hamiltonian formulations of thermostated
mechanics, three of them kinetic, and the other one configurational.
Though all four approaches ``work'' at equilibrium, their application
to many-body nonequilibrium simulations can fail to provide a proper
flow of heat.  All the Hamiltonian formulations considered here are
applied to the same prototypical two-temperature ``$\phi^4$''
model of a heat-conducting chain.  This model incorporates
nearest-neighbor Hooke's-Law interactions plus a quartic tethering potential.
Physically correct results, obtained with the isokinetic Gaussian and
Nos\'e-Hoover thermostats, are compared with two other Hamiltonian
results. The latter results, based on constrained Hamiltonian
thermostats, fail correctly to model the flow of heat.

\end{abstract}

\maketitle

\noindent

\vspace{0.4cm}

\section{Introduction}

Computational ``Thermostats'' arose as a means for controlling numerical
simulations of both equilibrium and nonequilibrium stationary states.
{\em Without} thermostats systems driven away from equilibrium heat up.
{\em With} thermostats the heat generated by irreversible processes can be
steadily removed, making it possible to simulate nonequilibrium steady
states\cite{b1,b2}. Because most Hamiltonian-based mechanics problems conserve
energy, novel {\em nonHamiltonian} ideas are typically required when thermostats
are to be included.
Nevertheless, several approaches to Hamiltonian thermostats have been
developed.  Here we consider various approaches pioneered by Bill Ashurst,
Carl Dettmann, Denis Evans, Wm. G. Hoover, Tom Leete, Gary Morriss,
Shuichi Nos\'e,  and Les Woodcock over a span of about 25
years\cite{b1,b2,b3,b4,b5,b6,b7,b8,b9,b10}.

The simplest thermostat type maintains a (nearly) constant kinetic energy by
``rescaling'' the velocities at the end of each computational timestep.  For
a more elegant, but equivalent, {\em continuous} approach to rescaling, see
the ``Gaussian isokinetic'' method described in Section IIb and illustrated
in Section VII.
For $\#$ Cartesian degrees of freedom the kinetic-theory relation,
$$
K = \sum^{\#}\frac{m\dot q^2}{2} = \sum^{\#}\frac{p^2}{2m} = 
\# \frac{kT_{\rm kin}}{2} ,
$$
defines the kinetic temperature $T_{\rm kin}$.  Thermostats based on
$T_{\rm kin}$ can be
applied to an entire system, or separately to two or more subsystems.  With
this {\em ad hoc} rescaling approach,
$$
p \rightarrow p_0\sqrt{(K/K_0)} \ ,
$$
there is no difficulty in maintaining several
temperatures in specified parts of a single nonequilibrium system, as did
Ashurst in his Ph D thesis research\cite{b11}.

Kinetic theory and the Gibbs-Boltzmann development of thermodynamics
based on ideal-gas thermometry certainly suggest that the kinetic temperature,
$T_{\rm kin}$,
is both the simplest and the most fundamental of the many possible
``temperature'' types.  There are alternatives.  Among the many possible
temperatures based on particle coordinates, rather than velocities, the
simplest---on which we focus here---is based on the mean-squared
forces, $\langle F^2 \rangle $.  This definition of a configurational
temperature is derived directly from the potential energy $\Phi $ and its
space derivatives:
$$
kT = \langle (\nabla \Phi)^2 \rangle /  \langle \nabla ^2 \Phi \rangle
= -\langle F^2 \rangle /  \langle \nabla F \rangle \ .
$$
This somewhat cumbersome relation appears as an aside in Landau and Lifshitz'
classic 1958 text\cite{b12}.  It was also recently rediscovered by Rugh,
and has subsequently been much discussed\cite{b13}.  This configurational
representation of temperature follows most simply from an integration by
parts, in the canonical ensemble:
$$
\int dq \nabla ^2\Phi = \nabla \Phi \ ; \ 
(-d/dq)  e^{-\Phi /kT} = (\nabla \Phi /kT) e^{-\Phi /kT} \longrightarrow
$$
$$
\int dq \nabla ^2\Phi e^{-\Phi /kT} =
(1/kT)\int dq (\nabla \Phi )^2e^{-\Phi /kT} .
$$

About ten years after Woodcock and Ashurst's introduction of the rescaling
isokinetic thermostat in the early 1970s, Hoover and Evans discovered that
the isokinetic velocity-rescaling equations of motion can be derived from
Gauss' Principle of Least Constraint\cite{b5}.  Fifteen years after this
discovery, Dettmann and Morriss\cite{b4} found a straightforward Hamiltonian approach
to these same motion equations.  Previously, Leete and Hoover had derived a
Hamiltonian formulation which likewise maintained the velocity-based kinetic
energy $K( \{ \dot q \} )$ constant\cite{b6,b8}.

A Hamiltonian generating the Gauss'-Principle isokinetic equations of motion was
not found until 1996\cite{b4}.  In the meantime Nos\'e had
discovered his canonical-ensemble thermostated dynamics, which is
related to two rather different Hamiltonians\cite{b7,b9}.   The
purpose of the present work is twofold.  First, we explore the relation
of the Hoover-Leete kinetic thermostat work to its successors.  A 2007 literature search
shows only a single reference to it\cite{b6}.  We also explore the
consequences of a ``Landau-Lifshitz'' configurational thermostat based on the
 canonical-ensemble
definition rediscovered by Rugh.

\section{The Nos\'e-Hoover Thermostat}

The Nos\'e-Hoover equations of motion\cite{b1,b9} are a standard approach to
carrying out canonical-ensemble dynamics at a particular temperature $T_0$.
This approach uses {\em feedback}, controlling the fluctuating kinetic energy
with a ``friction coefficient'' or ``control
variable'' $\zeta $.  The frictional feedback force is $-\zeta p$.   The time rate
of change of the friction coefficient is
based on the current value of the kinetic energy $K(t)$ relative to the
desired mean value $K_0(T_0)$.  The equations of motion also include a characteristic
time $\tau $ which determines the time-range over which the feedback acts:
$$
\{ \ \dot q = p/m \ ; \ \dot p = F - \zeta p \ \} \ ; \
\dot \zeta = [(K/K_0) - 1]/\tau ^2 \ .
$$
In most cases, including our $\phi ^4$ model studies described in
detail in Sections V-IX, a reasonable choice of $\tau $ corresponds to a typical
collision time or vibration time.  Such a choice typically
provides Gibbs' equilibrium canonical phase-space distribution with a
Gaussian distribution for the friction coefficient $\zeta $:
$$
f(q,p,\zeta ) = e^{-\Phi /kT}e^{-K/kT}e^{-\#\zeta ^2 \tau^2/2}
$$
Evidently the amplitude of the fluctuations of $\zeta $ varies as
$\sqrt{1/\#\tau ^2}$, where $\#$ is
the number of degrees of freedom.  Thus its effect on the dynamics vanishes
in the large-system limit.

{\em Nonequilibrium} situations are quite different.  By using two or more different
friction coefficients (or ``thermostat variables'') temperature differences
can be established, leading to heat flow.  Then the corresponding phase-space
distributions are no longer smooth and Gibbsian, but instead become
fractal, with the underlying phase-space trajectories satisfying the Second Law of
Thermodynamics infinitely more probable than those violating the Law\cite{b2}.  In
the equilibrium case there is a close connection of Nos\'e-Hoover mechanics
to Hamiltonian mechanics.  We describe this connection next.

\subsection{The Dettmann-Nos\'e Hamiltonian}

In 1984 Shuichi Nos\'e discovered a Hamiltonian consistent with
Gibbs' constant-temperature canonical ensemble,
$$
f(q,p) \propto \exp(-{\cal H}_{{\rm Nos\acute{e}}}/kT) \ .
$$
He was able to derive the Nos\'e-Hoover equations of motion described above from his
Hamiltonian by an artificial ``time scaling''.

In July of 1996 Carl Dettmann\cite{b7} discovered a simpler approach to the
Nos\'e-Hoover equations of motion. Dettmann's {\em vanishing} Hamiltonian is
$$
{\cal H}_{{\rm Dettmann}} \equiv s{\cal H}_{{\rm Nos\acute{e}}} =
\sum\frac{p^2}{2ms} +
s\left[  \ \Phi + \frac{p_s^2}{2M} + \#kT\ln s \ \right] \equiv 0 \ .
$$
Provided that Dettmann's Hamiltonian is set equal to the special value, zero, 
Nos\'e's time-scaling variable $s$ can be eliminated.  Then the Nos\'e-Hoover
equations of motion result {\em without the need for any time scaling}.  It needs
to be emphasized that this ${\cal H} = 0$ trick does {\em not} work for the
prototypical {\em nonequilibrium} case of a system with two different temperatures.
In a two-temperature system the two different variables $s_{\rm Hot/Cold}$ are
necessarily coupled and cannot both be eliminated, so that Dettmann's
${\cal H} = 0$ trick does not work.

There is a special case of Nos\'e-Hoover
dynamics that deserves special mention, the ``isokinetic'' case in which the
temperature is constant.  This corresponds to the choice $\tau \rightarrow 0$.
For isokinetic mechanics Dettmann and Morris discovered the special Hamiltonian
detailed next.

\subsection{The Dettmann-Morriss Isokinetic Hamiltonian}

Soon after Dettmann discovered a Hamiltonian route to the Nos\'e-Hoover motion
equations, Dettmann and Morriss\cite{b4} discovered a related approach
to isokinetic (constant temperature, provided that temperature is defined
by the kinetic energy) Hamiltonian dynamics:
$$
{\cal H}_{{\rm DM}} = Ke^{+\Phi /2K_0} - K_0e^{-\Phi /2K_0} \ \equiv 0 \ ; \
K \equiv \sum \frac{p^2}{2m} \ .
$$
As usual, the equations of motion follow by differentiation:
$$
\{ \ \dot q = \partial {\cal H}/\partial p = pe^{+\Phi /2K_0} \ ; \
\dot p = -\partial {\cal H}/\partial q = Fe^{-\Phi /2K_0} \ \} \ .
$$
The accelerations which follow, using the identity,
$$
K/K_0 = e^{-\Phi /K_0} \ ,
$$
are then exactly the same as those from the isokinetic
equations of motion given above:
$$
\{ \ \ddot q = (F/m) - p\sum (F\cdot \dot q/2K_0)e^{+\Phi /2K_0} \ = \
(F/m) - \zeta \dot q \ \} \ ; \ \zeta = \sum F \cdot \dot q/(2K_0) \ .
$$

\section{The Hoover-Leete Isokinetic Thermostat
in its Lagrangian and Hamiltonian Forms}

The Hamiltonians corresponding to the isokinetic Gauss'-Principle and
Nos\'e-Hoover velocity-based approaches are not
the only such means of thermostating equilibrium systems.  A straightforward
application of nonholonomic Lagrangian mechanics, as well as the familiar
Hamiltonian mechanics outlined in Leete's Master's thesis\cite{b6}, lead to
another type of isokinetic mechanics.  These two equivalent forms of
mechanics, both of which can be used to keep the {\it velocity}-based kinetic
energy $K(\dot q)$ constant, while allowing the momentum-based kinetic energy
$K(p)$ to vary, proceed by modifying the $\{ \dot q \}$, rather than the
$\{ \dot p \}$, equations.

To distinguish the two different kinetic energies we use the notation:
$$
K(\dot q) = \sum \frac{m\dot q^2}{2} \ ; \
K(p) = \sum \frac{p^2}{2m} \ .
$$
Let us begin with the Lagrangian case:
$$
{\cal L_{\rm HL}} \equiv K(\dot q) - \Phi + \lambda [K(\dot q)-K_0] \ ,
$$
where $K(\dot q)$ is the {\it velocity}-based kinetic energy,
$\sum \frac{m}{2}\dot q^2$,
and the Lagrange multiplier $\lambda  $ has the task of maintaining
$K(\dot q)$ at its initial value, $K_0$, as the motion proceeds.
Lagrange's equations of motion follow from the usual textbook differentiations
of the Lagrangian with respect to the velocities and coordinates:
$$
\{ \ p = \frac{\partial {\cal L_{\rm HL}}}{\partial \dot q} =
m\dot q(1+\lambda) \ ; \
\dot p = m\ddot q(1+\lambda) + m\dot q\dot \lambda =
\frac{\partial {\cal L_{\rm HL}}}{\partial q} \ \equiv F \ \} \ .
$$
Now multiply the $\dot p$ equation by $\dot q$ and sum:
$$
\sum m\ddot q(1+\lambda)\dot q + \sum m\dot q\dot \lambda \dot q = 0 + 
2K_0\dot \lambda = \sum F\cdot \dot q \equiv -\dot \Phi \ .
$$
The value of the Lagrange multiplier $\lambda $ follows:
$$
\lambda = \frac{(\Phi _0 - \Phi)}{2K_0} = \frac{\sqrt{4K(p)K_0} - K_0}{2K_0} 
= \frac{2K_0(1+\lambda ) - 2K_0}{2K_0} = \lambda \ ,
$$
where the last expression, which completes the identity, follows from the
Hamiltonian given just below.

With Leete's help, Hoover discovered, in 1979, that the {\em velocity}-based
kinetic energy $K(\dot q)$ can alternatively be kept constant by using the
(constant in time) Hoover-Leete Hamiltonian:
$$
{\cal H_{\rm HL}} \equiv \sum \dot qp - {\cal L_{\rm HL}} =
\sqrt{4K_0K(p)} + \Phi - K_0 \ = \ {\cal H}_0 =
\Phi _0 + K_0,
$$
where $\Phi $ is again the usual potential energy and $\Phi _0$ is its initial
value.
 
The constancy of the kinetic energy is easy to see.  The equations of motion,
$$
\{ \ \dot q = \frac{\partial {\cal H}}{\partial p} \ ; \
\dot p = -\frac{\partial {\cal H}}{\partial q} \ \} \  \longrightarrow
\{ \ \dot q = \frac{p}{m}\sqrt{\frac{K_0}{K(p)}} \ ; \ \dot p = F(q) \ \} \ ,
$$
imply that the velocity-based kinetic energy does not vary:
$$
K(\dot q) = \frac{m}{2}\sum \dot q^2 =
\sum \frac{p^2}{2m}\left[\frac{K_0}{K(p)}\right] \equiv K_0 \ .
$$
At the same time, except in the equilibrium case with a large number of
degrees of freedom, there is no guarantee that the momentum-based kinetic energy $K(p)$ is 
similar in magnitude to $K_0$.
But, provided that $\{ \ m\dot q = p \ \} \rightarrow K(p) = K(\dot q) = K_0$
initially, Leete's equations of motion do match the Gaussian isokinetic
ones to second order,
$$
\{ \ \ddot q = \frac{F}{m} - \zeta \dot q \ \} \ ; \
\zeta = \frac{\sum F \cdot \dot q}{2K_0} \ ,
$$
through the second derivatives, $\{ \ \ddot q \ \}$, but differ
in the {\em third} derivatives, where the meansquared constraint force
in the isokinetic case is less than the Hoover-Leete analog, in accordance with Gauss'
Principle of Least Constraint\cite{b1,b5}.

The constraint of constant temperature in Hamiltonian mechanics contradicts
the thermodynamic notion that the energy and temperature cannot be varied
independently in a system of fixed composition and volume.  Both $E$ and $T$
are {\em constants} of the motion, using Leete's approach.  Provided that
the initial conditions are wisely chosen, with $E$ and $T$ corresponding to
the {\em same} thermodynamic state, this approach {\em can} certainly be
used to determine equilibrium properties.  We will see, in Section VIII,
what the (rather strange) consequences of this thermostat are away from
equilibrium.

\section{Landau-Lifshitz' Configurational Thermostat}

A (much) more complicated Hamiltonian thermostat, conserving the force-based
{\em configurational} temperature $T_{\rm con}$, can be based on
straightforward (though quite tedious) holonomic Hamiltonian mechanics.
A proper configurationally thermostated Hamiltonian conserves not only
the Hamiltonian but also the configurational temperature,
$$
kT_{\rm con} \equiv \sum _N F_i^2/\sum _N \nabla_i^2{\cal H} \ .
$$
Here $k$ is Boltzmann's constant, which we set equal to unity in the
numerical work.  A molecular dynamics simulation based on this definition
of temperature follows standard Hamiltonian mechanics, as the temperature
constraint is just a (complicated) holonomic (coordinates only) constraint.
The simplest procedure begins with the system Lagrangian, augmented with
a Lagrange multiplier $\lambda $ which constrains
$\{ \  T_{\rm con},\dot T_{\rm con},\ddot T_{\rm con} \ \}$.  The
corresponding Hamiltonian has the form,
$$
{\cal H} = \Phi (\{ q\} ) + K (\{ p\} ) + \lambda ( T_{\rm con} - T_0) \ .
$$

For success, the initial conditions have to be chosen to correspond to
both the desired temperature, $T_{\rm con} = T_0$, and to the condition
$\dot T_{\rm con} = 0$.  Then, to begin the analytic work, differentiate the
temperature equation, $T_{\rm con}(\{ q \}) = T_0$, twice with
respect to time.  The two time differentiations, using the chain rule, 
give first $\{ \dot q\} $ and then $\{ \ddot q\} $.  By substituting
the constrained equations of motion:
$$
\{ \ \ddot q_i = \ddot \delta _i = \dot p_i =
F_i(\{q\}) - \lambda \nabla_iT_{\rm con} \ \} \ ,
$$
for the $\{ \ddot q \}$, we obtain the Lagrange multiplier $\lambda $.  For
further details of this calculation, see Section IX.  The
numerical work can be checked by noting that both $T = T_{\rm con}$ and ${\cal H}$
are constants of the motion when the calculation is error free.

\section{Aoki and Kusnezov's $\phi^4$ Model for Heat Conduction}

The $\phi^4$ model gets its name from the functional form of a quartic ``tethering
potential'', $\Phi_{\rm Teth}$, which links each particle to a fixed
lattice site with a cubic restoring force:
$$
\Phi_{\rm Teth} \equiv \sum _N\left(\frac{\kappa \delta ^4}{4}\right)
\ \rightarrow F_{\rm Teth} = -\kappa \delta ^3 \ .
$$
In the pedagogical simulations which follow we will choose the tethering force constant
equal to unity, $\kappa = 1$.  Nearest-neighbor particle pair
interactions in the $\phi^4$ model are governed by Hooke's Law:
$$
\Phi_{\rm Pair} = \sum_{\rm NN}\left(\mu /2\right)(|r|-d)^2 \ .
$$

Aoki and Kusnezov have carried out a variety of pedagogical heatflow
simulations for this model in both one and three space
dimensions\cite{b14,b15}.
The $\phi^4$ model has a finite Fourier conductivity in one dimension.  
The numerical work carried out by Aoki and Kusnezov established
the temperature dependence of the thermal conductivity, $3/T^{4/3}$.

For simplicity in the one-dimensional work which we carry out here, we always
choose the nearest-neighbor separation $d$ equal to unity.  We also choose
the strength of the Hooke's-Law interaction $\mu $ equal to
unity.  As was abundantly demonstrated by Aoki and Kusnezov, the combination
of a site-based tethering potential with a Hooke's-Law pair
potential provides the usual Fourier conductivity, with the heat flux
proportional to the (sufficiently small) temperature gradient.  In what
follows we apply four different thermostat constraints to this potential
model, using four different dynamical approaches:
Nos\'e-Hoover, Gaussian isokinetic, Hoover-Leete isokinetic, and
Landau-Lifshitz isoconfigurational.

A numerical solution of the heat flow equation for the one-dimensional chain,
$$
\dot T = \nabla (3T^{-4/3}\nabla T) \pm \alpha T \ ,
$$
can be obtained by Runge-Kutta integration, with a rescaling of the
temperatures within the reservoir regions at the end of each timestep
accounting for the source and sink terms $\pm \alpha T$ in the flow equation.
Figure 1 shows a fully-converged temperature profile obtained in this way,
using hot and cold temperatures of 0.26 and 0.24, with 400 mesh points.
The spatial gradient operations were replaced by finite-difference
approximations:
$$
(\nabla T)_i = (T_{i+1} - T_{i-1})/2 \ ,
$$
with appropriate subscript changes accounting for the periodic boundary 
conditions.

\section{Results with the Nos\'e-Hoover Thermostat}

With a periodic system composed of four parts, ``Hot'', ``Newton$_1$'',
``Cold'', and ``Newton$_2$'', each part containing $N/4$ particles, the
equations
of motion are
$$
\{ \ m\ddot q = F - \zeta _{\rm Hot}m\dot q \ \} \ ; \ \dot \zeta _{\rm Hot} = 
[(K/K_{\rm Hot}) - 1]/\tau ^2 \ ;
$$
$$
\{ \ m\ddot q = F - \zeta _{\rm Cold}m\dot q \ \} \ ; \ \dot \zeta _{\rm Cold} = 
[(K/K_{\rm Cold}) - 1]/\tau ^2 \ ;
$$
in the hot and cold regions, and
$$
\{ \ m\ddot q = F \ \} 
$$
in the two Newtonian regions.  The $\phi ^4$ force $F_i$ depends upon the three
coordinates
$\{ x_{i-1},x_i,x_{i+1}\}$ and also includes the tethering force, 
$F_{\rm Teth} = -\kappa (x_i - x_0)^3$, where $x_0$ is the lattice site for
the $i$th particle in the perfect lattice.  The initial velocities and
displacements were selected randomly within ranges:
$$
-\Delta _v < \{ v\} < +\Delta _v \ ; \ -\Delta _q < \{ \delta q\} < +\Delta _q
\ ,
$$
with $\Delta _v$ or $\Delta _q$ chosen to reproduce the desired stationary
reservoir temperatures, 0.26 and 0.24. The {\em average} temperatures, both
kinetic and configurational, from the last half of a 400-particle,
40-million timestep simulation,
$$
0 < t < 100,000 ({\rm equilibration}) \ ; \ 
100,000 < t < 200,000 ({\rm averaging}) \ ,
$$
are shown in Figure 2, along with a comparison solution of the macroscopic
heat flow equation.  The agreement is quite good, as would be expected from
Aoki and Kusnezov's work.

\section{Results with the Gaussian Isokinetic Thermostat}

Simulations with two {\em Gaussian isokinetic thermostats}, with the same
timestep, $dt = 0.005$, and the same equilibration and averaging intervals,
produced similar results, maintaining the initial kinetic temperatures
throughout.  The averaged profiles, along with a corresponding continuum
solution, are shown in Figure 3.  Again the agreement, for both the
kinetic and the configurational temperatures, is perfectly satisfactory.

The instantaneous (at time $t$) values of the total heat added (from time
0 to time $t$) to the ``hot'' reservoir as well as that removed from the
``cold'' reservoir are shown in Figure 4. The sum of the two necessarily
oscillates about zero as $t$ increases.
Straight lines drawn through the data provide an estimate for the Fourier
conductivity in agreement with the results of Aoki and Kusnezov's work.
The two conventional thermostat types, Nos\'e-Hoover and Gaussian isokinetic,
show that either method can be used to simulate the simple two-temperature
heat flow problem.

\section{Results with the Hoover-Leete Hamiltonian Isokinetic Thermostat}

Results with the Hamiltonian isokinetic thermostats:
$$
\{ \ \dot q_{\rm Hot/Cold} =
\frac{p}{m}\sqrt{\frac{K_{\rm Hot/Cold}}{K(p)}} \ ; \ \dot p = F(q) \ \} \ ,
$$
where the fixed hot and cold kinetic temperatures
$\langle m\dot q^2/2k \rangle _{\rm Hot/Cold}$ are fixed by the initial
conditions, but come to differ from the unconstrained momentum-based
temperatures
$\langle p^2/2mk \rangle _{\rm Hot/Cold}$.  Unlike the Nos\'e-Hoover and
Gaussian isokinetic profiles, the Hamiltonian-based profiles show {\em no
temperature gradients at all}.  See Figure 5.  Instead there are
discontinuities between
the fixed kinetic temperatures and the adjacent constant values of the
unconstrained Newtonian regions.  In the latter regions the kinetic and
configurational temperatures equilibrate, and match the unconstrained
configurational temperature of the thermostated regions.  In the
thermostated regions there is no such equilibration.

Evidently there are no heat fluxes in the system either.  By constraining
not only the energy (through the constant Hamiltonian driving the motion)
but also the kinetic temperatures, the system is evidently prevented
from supporting heat flow.  The Hamiltonian nature of the motion equations
also prevents the formation of the multifractal phase-space distributions
associated with nonequilibrium stationary states\cite{b1,b2}.  A more
detailed analysis of the reasons for the failure of this Hamiltonian 
approach could perhaps be based on a detailed analysis of the microscopic
equations describing heat flux\cite{b1}.

\section{Results with the Landau-Lifshitz Hamiltonian Isoconfigurational Thermostat}

For the special $\phi^4$ model considered here, with $d=1$, the force $F_i$ is
$$
F_i = -\kappa \delta _i^3 + \mu [\delta_{i+1} - 2\delta_i + \delta_{1-1}] \ .
$$
The {\em pair-potential} contribution to $\nabla_i ^2{\cal H}$ is $2\mu $, for
the two nearest-neighbor interactions,
$$
\frac{\partial ^2}{\partial x_i^2}(\mu /2)(|x_i - x_{i\pm 1}| - 1)^2
= \mu \ .
$$
The {\em tethering} potential contribution is
$3\kappa \delta_i^2 = \nabla _i^2(\kappa \delta _i^4/4)$:
$$
\sum _N \nabla^2{\cal H} = 2N\mu + \sum _N3\kappa \delta^2 \ .
$$
 
The time derivative of the configurational temperature is a quotient of
products of single-particle sums:
$$
\dot T = \frac{2\sum F\dot F}{\sum(2\mu+3\kappa \delta ^2)}
 - \frac{6\sum F^2\sum \kappa v\delta }{[\sum(2\mu +3\kappa \delta ^2)]^2} \ ,
$$
where
$$
\dot F_i = \mu [v_{i+1} - 2v_i + v_{i-1}] - 3\kappa \delta _i^2v_i \ .
$$
Another time differentiation gives the second derivative:
$$
\ddot T_{\rm Newton} =
\frac{2\sum F\ddot F+2\sum \dot F^2}{\sum(2\mu +3\kappa \delta ^2)} \
+ \
\frac{72\sum F^2[\sum \kappa v\delta ]^2 }{[\sum(2\mu +3\kappa \delta ^2)]^3}
\ -
$$
$$
\frac{24\sum F\dot F\sum \kappa v\delta +6\sum F^2(\kappa\sum [a\delta + \sum
  v^2])}{[\sum(2\mu +3\kappa \delta ^2)]^2} \ .
$$
where
$$
\ddot F_i = \mu[a_{i+1} - 2a_i + a_{i-1}] -
\kappa [3 \delta _i^2a_i + 6\delta _iv_i^2] \ .
$$
The {\em second} derivative, $\ddot T$, introduces the Lagrange multiplier
through the equations of motion:
$$
\{ \ a_i = \dot v_i = F_i - \lambda \nabla _iT \ \} \ .
$$
Though complicated, the resulting equation for $\lambda $ is linear, and gives
an explicit expression for the Lagrange multiplier that stabilizes $T$:
$$
\lambda = \frac{\ddot T_{\rm Newton}}
{\sum \nabla _iT\left([2\mu (F_{i+1}-2F_i+F_{i-1}) - 6\kappa \delta ^2 _iF_i]
- [\sum F^2\sum 6\kappa F_i\delta _i]\right)} \ .
$$
Because $T$ cannot change, both $T=T_0$ and $\dot T=0$
need to be properly specified in the initial conditions.  A convenient way to
do this is to choose the velocities equal to zero so that  $\dot T \equiv 0$.

A typical profile using two configurational thermostats is shown in Figure
6.  Just as in the Hoover-Leete isokinetic case there is no equilibration of
the constrained temperature (here configurational).  The unconstrained kinetic
temperature equilibrates throughout the system to a value dependent upon the
initial conditions.  Evidently the Landau-Lifshitz constrained configurational
thermostat is quite useless for nonequilibrium work.

\section{Conclusions}

The results obtained with the Gaussian isokinetic and Nose-Hoover thermostats
are quite consistent with the past history of their use over the last quarter
century\cite{b1,b2,b6,b7}.  For these familiar thermostats there was no problem
in reaching a nonequilibrium steady state with temperature profiles and heat
fluxes very similar to those expected from macroscopic hydrodynamics using
Fourier's law.  On the other hand, the results obtained with {\em purely
Hamiltonian} mechanics, using either the Hoover-Leete constrained isokinetic
or the Landau-Lifshitz constrained configurational thermostats were both
unexpected and thought-provoking.  Both Hamiltonian approaches stably
constrained their target temperatures in the two reservoir regions.
Nevertheless temperature differences led to no net heat
flow and the computed temperature profiles bore no resemblance to the
predictions of hydrodynamics.  The kinetic and configurational temperatures,
though {\em unequal} in the constrained reservoirs,
$$
T_{\rm kin} \neq T_{\rm con} \ [{\rm reservoirs}] \ ,
$$
{\em were} equal (to yet a third nearly-constant temperature) in the unconstrained
bulk Newtonian regions:
$$
T_{\rm kin} = T_{\rm con} \ [{\rm bulk}] \ .
$$
These results emphasize the fact that {\em nonHamiltonian} dynamics is absolutely
necessary to the realistic modeling of nonequilibrium systems.  A direct
proof/understanding of the failure of the Hamiltonian systems to show a
nonvanishing heat flux would be most welcome.

\section{Acknowledgments}
This work was first presented at a Mexico City workshop organized by
Francisco Uribe at Universidad Aut\'onoma Metropolitana-Iztapalapa in January,
2007.  Subsequently, a shorter version was presented at the Schr\"odinger
Instutite workshop, in Wien, honoring the 65th birthday of Herr Professor
Doktor Harald Posch.  We are specially grateful to
Professors Uribe, Posch, and Dellago for making this work possible.

\pagebreak

\noindent
{\bf Figure Captions}

\noindent
1.  Temperature profile according to Fourier's law, with mean reservoir
temperatures of $0.26 \ (-200 < x < -100)$ and $0.24 \ (0 < x < 100)$.

\noindent
2.  Temperature profile for a conventional two-temperature Nos\'e-Hoover
simulation, as described in the text.

\noindent
3.  Temperature profile for a conventional two-temperature Gaussian
isokinetic simulation, as described in the text.

\noindent
4.  Summed-up values of the heat transferred to the hot and cold Gaussian
isokinetic reservoirs for the simulation of Figure 3.

\noindent
5.  Temperature profiles (kinetic and configurational) for a simulation
using constrained Hoover-Leete isokinetic reservoirs.  The timestep is
$dt = 0.005$, with a total elapsed time of $10^6$ divided equally between
equilibration and averaging portions.

\noindent
6.  Temperature profiles (kinetic and configurational) for a simulation
using constrained Landau-Lifshitz isoconfigurational reservoirs.  The timestep is
$dt = 0.002$, with a total elapsed time of $400,000$ divided equally between
equilibration and averaging portions.

\pagebreak


\begin{references}
 

\bibitem{b1}  Wm. G. Hoover, {\em Computational Statistical Mechanics}
              (Elsevier, New York, 1991).

\bibitem{b2}  Wm. G. Hoover, {\em Computer Simulation, Time Reversiblity,
              and Chaos} (World Scientific Publishing, Singapore, 1999 and
              2001).

\bibitem{b3}  W. T. Ashurst and W. G. Hoover, ``Argon Shear Viscosity
              {\em via} a Lennard-Jones Potential with Equilibrium and
              Nonequilibrium Molecular Dynamics'', Physical Review
              Letters {\bf 31}, 206-208 (1973).

\bibitem{b4}  C. P. Dettmann and G. P. Morriss, ``Hamiltonian Formulation of
              the Gaussian Isokinetic Thermostat'', Physical Review E {\bf 54},
              2495-2500 (1996).

\bibitem{b5}  D. J. Evans, W. G. Hoover, B. H. Failor, and B. Moran.
              ``Nonequilibrium Molecular Dynamics {\em via} Gauss' Principle
              of Least Constraint'', Physical Review A {\bf 28}, 1016-1021
              (1983).

\bibitem{b6}  W. G. Hoover, ``Atomistic Nonequilibrium Computer Simulations'',
              Physica {\bf 118A}, 111-122 (1983).

\bibitem{b7}  W. G. Hoover, ``M\'ecanique de Non\'equilibre \`a la
              Californienne'', Physica {\bf 240A}, 1-11 (1997).

\bibitem{b8}  T. M. Leete, ``The Hamiltonian Dynamics of Constrained Lagrangian
              Systems'' [M S Thesis, West Virginia University, 1979].

\bibitem{b9}  S. Nos\'e, ``Constant Temperature Molecular Dynamics Methods'',
              Progress of Theoretical Physics Supplement {\bf 103} 
              ({\em Molecular Dynamics Simulations}, S. Nos\'e, Editor),
              1-46 (1991).

\bibitem{b10} L. V. Woodcock, ``Isothermal Molecular Dynamics
              Calculations for Liquid Salts'', Chemical Physics
              Letters {\bf 10}, 257-261 (1971).

\bibitem{b11} W. T. Ashurst, ``Dense Fluid Shear Viscosity and Thermal
              Conductivity {\em via} Nonequilibrium Molecular Dynamics''
              [Ph D Dissertation, University of California at
              Davis/Livermore, 1974]).

\bibitem{b12} L. D. Landau and E. M. Lifshitz, ``Statistical Physics'',
              equation 33.14 (McGraw-Hill, New Jersey, 1958).

\bibitem{b13} J. G. Powles, G. Rickayzen, and D. M. Heyes, ``Temperatures:
              Old, New and Middle-Aged'',  Molecular Physics {\bf 103},
              1361-1373 (2005).

\bibitem{b14} K. Aoki and D. Kusnezov, ``Nonequilibrium Steady States and
              Transport in the Classical Lattice $\phi ^4$ Theory'', Physics
              Letters B {\bf 477}, 348-354 (2000).  Conductivity is
              $2.72T^{-1.38}$.

\bibitem{b15} K. Aoki and D. Kusnezov, ``Violations of Local Equilibrium and
              Linear Response in Classical Lattice Systems'', Physics Letters
              A {\bf 309}, 377-381 (2003). Conductivity is
              $2.8T^{-1.32}$.



\end{references}
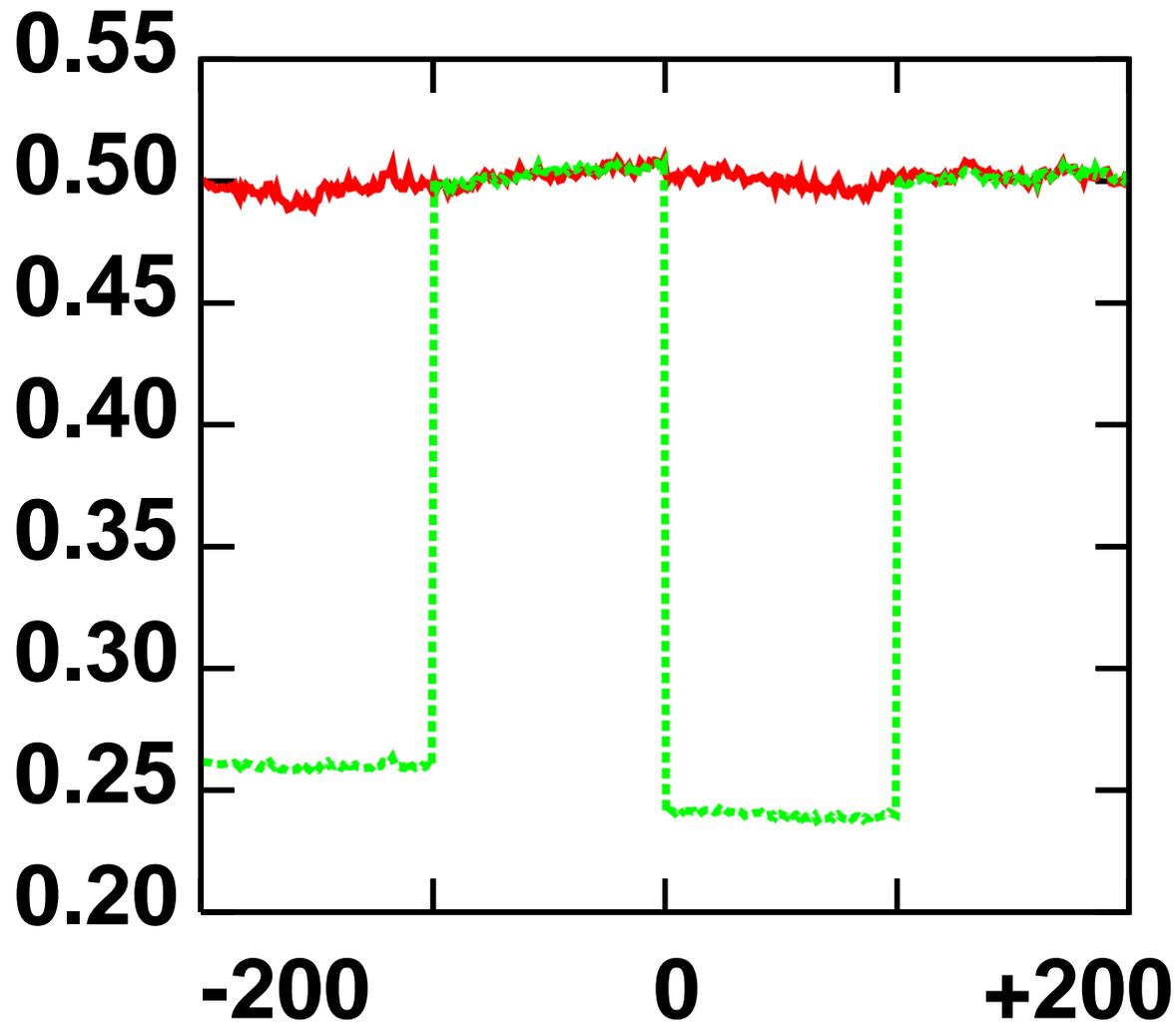
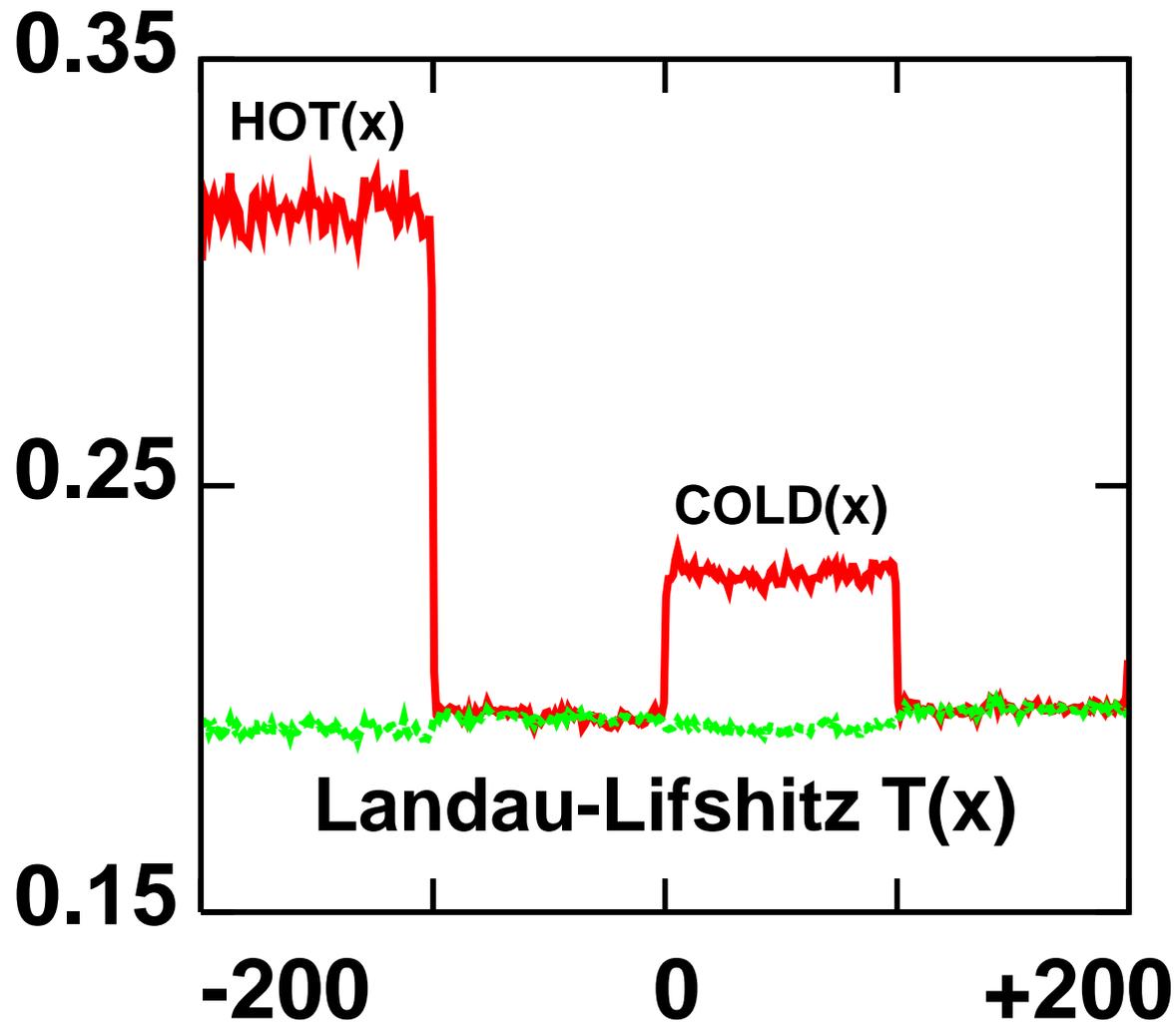
\end{document}